\documentstyle[11pt,newpasp,twoside,epsf]{article}

\gdef\cl1{RXJ\,0848+4453}
\gdef\zform{z_{\rm form}}

\begin{document}
\title{Formation and Evolution of Galaxies in Clusters}
 \author{Pieter G. van Dokkum}
\affil{California Institute of Technology, MS 105-24, Pasadena, CA 91125}

\begin{abstract}

Elliptical and S0 galaxies dominate the galaxy population in nearby
rich clusters such as Coma. Studies of the evolution of
the colors, mass-to-light ratios, and line indices of early-type
galaxies indicate
that they have been a highly homogeneous, slowly
evolving population over the last $\sim 65$\,\% of the age of the Universe.
On the other hand, recent evidence suggests that many early-type galaxies
in clusters have been transformed from spiral galaxies since $z \sim 1$.
Arguably the most spectacular evidence for such transformations
is the incidence of red merger systems in
several high redshift clusters.
Due to this morphological evolution
the sample of early-type galaxies at high redshift is only a subsample
of the sample of early-type galaxies at low redshift. 
This ``progenitor
bias'' results in an overestimate of the mean formation redshift
if simple models without morphological transformations are used.
Models
which incorporate morphological evolution explicitly can bring the
homogeneity, slow evolution, and morphological transformations
into agreement. The modeling shows that the corrected mean formation
redshift of the stars in early-type galaxies may be as low as
$z\approx 2$ in a $\Lambda$ dominated Universe.

\end{abstract}

\section{Introduction}

The galaxy population in rich clusters is dominated by early-type
galaxies (S0 and elliptical galaxies).  The study of these objects
gives insight in the formation of the most massive disk- and spheroidal
galaxies in the Universe, and in the processes governing star
formation at early times. Furthermore, cluster galaxies provide
critical tests of the hierarchical paradigm for galaxy formation. In
currently popular semi-analytical galaxy formation models in a CDM
Universe the descendants of the Ly-break population are massive
galaxies in groups and clusters (Baugh et al.\ 1998).
These models also predict that significant differences
should exist between early-type galaxies in clusters and those in the
general field (Kauffmann 1996).

It has been known for a long time that early-type galaxies in clusters
form a very homogeneous population: at a given
luminosity, they show a very small scatter in their colors,
$M/L$ ratios, and line indices (e.g., Bower et al.\ 1992).
The simplest interpretation of this high degree of homogeneity
is a small spread in age, although it has been argued that a larger
age spread could be ``masked'' by correlated metallicity
variations (e.g., Trager et al.\ 2000)\footnote{Note that this
interpretation requires that we observe early-type galaxies at a special time,
when age and metallicity variations exactly cancel.}.
Determining the {\em mean} age of nearby
early-type galaxies has proven
to be a formidable challenge.
The main reason is the well known
age-metallicity degeneracy in fitting early-type galaxy
spectra (e.g., Worthey 1994). Furthermore, the observed abundance ratios
of early-type galaxies cannot be reproduced with simple stellar population
synthesis models, which makes absolute determinations of age and
metallicity even more uncertain. It is therefore not surprising
that most of our understanding
of the formation and evolution of cluster early-type galaxies has come from
studies of clusters at large lookback times.

Since the seminal work by Butcher \& Oemler (1978) on the colors
of galaxies in two distant clusters
this field has witnessed
great progress: measurements of redshifts, morphologies, colors, $M/L$ ratios,
and line indices of cluster galaxies currently
span $\sim 65$\,\% of the age of the Universe.
Examples of successful ongoing programs are the MORPHS
collaboration (Smail et al.\ 1997), who obtained deep HST images of
the cores of $\sim 10$ clusters at $0.3<z<0.5$, the CNOC group (Yee et
al.\ 1996), who obtained extensive wide field spectroscopy and imaging
of X-ray selected clusters at $0.2<z<0.5$, the work by Lubin, Postman,
\& Oke on optically selected clusters at $z \sim 0.8$, and our wide
field HST imaging and extensive spectroscopy of X-ray
clusters\footnote{Possibly dubbed AWACS, for A Wide Angle Cluster
Survey.}.

As in many other fields, Hubble Space Telescope imaging and
spectroscopy with large ground-based telescopes have been instrumental
in this advance. An additional factor has been the success of surveys
of increasing sophistication to find ever more distant clusters.  The
progress in this area that was demonstrated at the meeting is
encouraging, and should lead to a better understanding of the
interplay between the selection of clusters and the derived evolution
of the galaxies within them (see, e.g., the review by Marc
Postman).

\section{Evolution of early-type galaxies}

Studies of the evolution of early-type galaxies in clusters
are in remarkable agreement. Studies of their colors (e.g., Ellis et al.\ 1997,
Stanford et al.\ 1998), $M/L$ ratios (van Dokkum \& Franx 1996,
Bender et al.\ 1998, van Dokkum et al.\ 1998, Kelson et al.\ 2000),
and line indices (Bender et al.\ 1998, Kelson et al.\ 2001) 
show that they remain a very homogeneous population and
evolve only slowly all the way from $z=0$ to $z \sim 1$.

\subsection{Evolution of the Fundamental Plane}

The strongest constraints on the mean star formation epoch
have come from the evolution of the
Fundamental Plane (FP) relation. The Fundamental Plane
(Djorgovski \& Davis 1987) is a relation between the effective radius
$r_e$, effective surface brightness $\mu_e$, and central velocity
dispersion $\sigma$, such that $r_e\mu_e^{0.8} \propto \sigma^{1.25}$
in the $B$ band. The implication of the existence of the FP is
that $M/L$ ratios of galaxies correlate strongly with their structural
parameters: $M/L \propto r_e^{0.2} \sigma^{0.4} \propto M^{0.2}$
(Faber et al.\ 1987). The power of the FP lies in this relation to
$M/L$ ratios, and its usefulness for galaxy evolution studies stems
from its small scatter and the fact that it applies to both
elliptical and S0 galaxies.

\begin{figure}[t]
\plotone{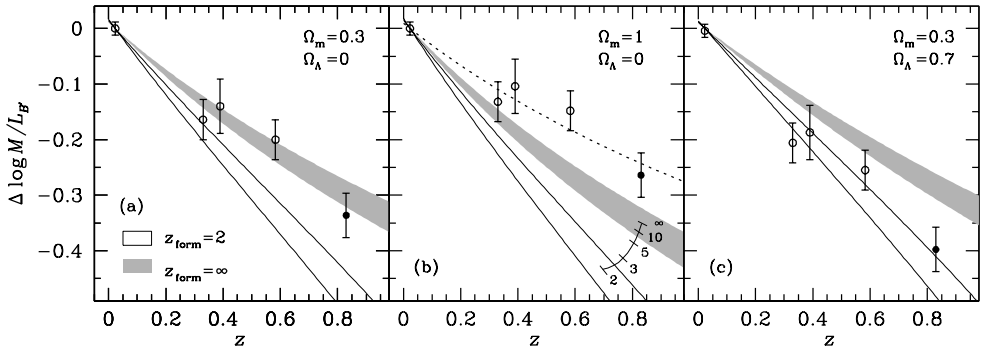}
\caption{Evolution of the mean $M/L_B$ ratio of early-type galaxies,
as determined from the FP relation. The observed evolution is slow,
indicating an early formation of the stars.}
\end{figure}

The $M/L$ ratios of galaxies are expected to evolve because the
luminosity of their stellar populations decreases as they
age (``passive evolution''). The {\em rate} of evolution depends
on the time that has elapsed since the population was formed:
the light of young stellar populations is dominated by massive
stars which have a short life time on the main sequence, whereas
the light of old stellar populations is dominated by long lived,
low mass stars.  In the rest frame $B$ band, the
expected evolution is $M/L \propto (t-t_{\rm form})^{0.91}$
for a Salpeter IMF;
the coefficient depends on the passband and the IMF, but is only
weakly dependent on the metallicity. As a result, the rate of evolution
of the intercept of the FP gives a strong constraint on the mean
stellar age of early-type galaxies.

The measured evolution of $M/L_B$ to $z=0.83$ is shown in Fig.~1,
from van Dokkum et al.\ (1998). The evolution is
surprisingly low, $\ln M/L_B \propto -z$, indicating stellar
formation redshifts of $z>2.8$ for $\Omega_m = 0.3$,
$\Omega_{\Lambda}=0$, and a Salpeter IMF.
Studies of the evolution of colors (e.g.,
Stanford et al.\ 1998) and line indices (Bender et al.\ 1998;
Kelson et al.\ 2001) have yielded very similar results. In general,
color evolution (effectively the difference between
the luminosity evolution in each of two passbands)
can not be measured to the same precision as
evolution in $M/L$ ratios.


It is difficult, but not impossible, to extend the FP measurements to
even higher redshift. The practical limit probably
lies around $z \sim 1.3$:
12.5\,hr Keck spectra of ``Extremely
Red Objects'' in the cluster \cl1\ at $z=1.27$ 
are just sufficient to measure velocity dispersions
(van Dokkum \& Stanford, in prep).

\subsection{Evolution of the scatter in the color-magnitude
relation}

The color-magnitude (CM) relation provides important additional
constraints on the star formation epoch of early-type galaxies.
Because spectroscopy is not required it is relatively
straightforward to obtain large samples, enabling studies of the
evolution of the scatter and slope of the relation as well as
its zeropoint.

The scatter is of particular interest, because it measures
the spread in stellar age among early-type galaxies. The rest frame
$U-V$ color evolution of
a stellar population can be described by $L_V/L_U \propto
(t-t_{\rm form})^{\kappa_U - \kappa_V}$, with $\kappa_U \approx 1.08$
and $\kappa_V \approx 0.81$.
It can be shown that the scatter in the CM relation at any time $t$
is proportional to the
scatter in luminosity weighted age
divided by the mean age (e.g., van Dokkum et al.\ 2000).

The observed evolution of the scatter in the CM relation
is shown in Fig.\ 2.
Ground based data are from Bower et al.\ (1992) and
Terlevich et al.\ (2001)
for Coma and from Stanford et al.\ (1998) for high redshift clusters.
HST measurements have smaller errorbars, and are from Ellis et al.\
(1997) and van Dokkum et al.\ (1998, 2000, 2001).
The scatter remains very small all the way from $z=0$ to $z=1.3$.
This result is quite surprising, because the mean age of galaxies
should be {\em at least} a factor 2 smaller at $z=1$. Therefore,
if the scatter in the CM relation at $z=0$ is caused by age
variations, one might expect the scatter to increase by at least
a factor 2 from $z=0$ to $z=1$. Following a similar
line of reasoning it has been argued that the scatter in the CM
relation of nearby clusters is mainly due to metallicity variations,
and that the formation of early-type galaxies is even more synchronized
than implied by the tight scaling relations observed at low redshift
(e.g., Stanford et al.\ 1998). Indeed, when taken at face value, the
scatter observed at $z=1.27$ implies a spread in age of only $\sim 5$\,\% 
at the present epoch!

\begin{figure}
\plotone[-78 216 663 619]{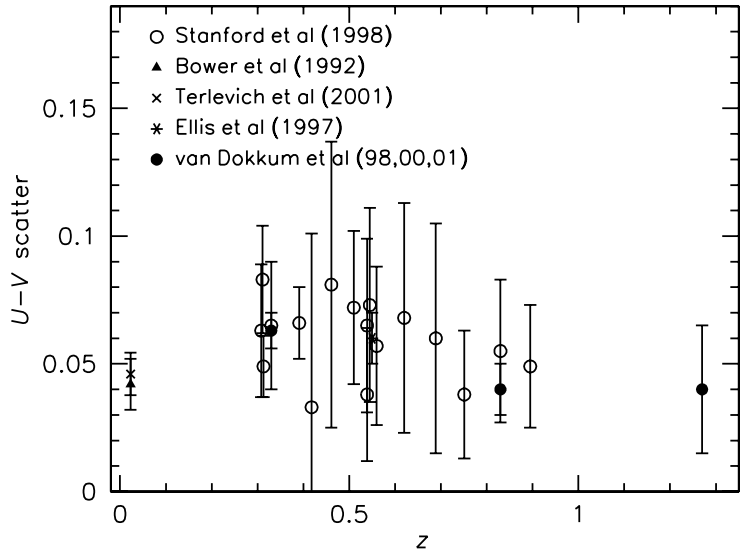}
\caption{Evolution of the scatter in the color-magnitude relation
with redshift, from literature data (see text). The scatter shows
little evolution, implying that the scatter in age is small at
all times.}
\end{figure}

In summary, studies of
the observed evolution of early-type galaxies indicate
that they have been a highly homogeneous, slowly evolving
population over at least the
latter $\sim 65$\,\% of the age of the Universe. To satisfy the tightest
observational constraints it appears that early-type galaxies
would have to have formed at very high redshift ($z>3$) in a very short
time ($\leq 500$\,Myr).

\section{Assembly time of cluster galaxies}

\subsection{The Butcher-Oemler effect}

It has been known for a long time that something must be amiss
with the simple picture of early formation presented above.
The earliest evidence for significant recent evolution in
cluster environments was the discovery of the Butcher-Oemler
effect: the increase with redshift of the fraction of blue galaxies in
clusters (Butcher \& Oemler 1978, 1984). Figure 3 is a compilation
showing the
evolution of the blue fraction with redshift, with data
from Butcher \& Oemler
(1984), Smail et al.\ (1998), Fabricant et al.\ (1991),
van Dokkum et al.\ (2000), and Ellingson et al.\ (2001).

\begin{figure}[h]
\plotone[-61 225 646 636]{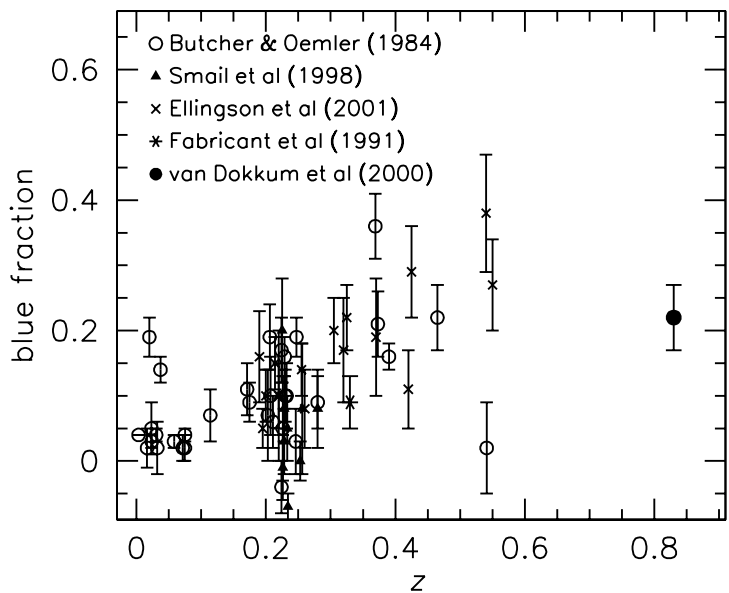}
\caption{Evolution of the blue galaxy fraction in 
clusters. The trend has significant scatter.}
\end{figure}

There is clearly a trend in Fig.\ 3, albeit with a large scatter.
Much of the work on
cluster galaxies in subsequent decades -- too much to
do justice in this short review -- was aimed at understanding
the nature of these blue galaxies. It has become clear that most
of the blue galaxies are not early-type galaxies but low mass spirals
and irregulars (e.g., Smail et al.\ 1997). Some show ongoing star formation;
others are currently not forming stars but have enhanced
Balmer absorption lines indicating a recent star burst
(e.g., Dressler \& Gunn 1983).
Among the more persistent
ideas is that the Butcher-Oemler effect is driven by infall of
blue, late-type galaxies from the field, which subsequently lose their fuel
for star formation in interactions with other galaxies and/or
the hot X-ray gas (e.g., Abraham et al.\ 1996, Ellingson
et al.\ 2001). In this picture,
the redshift dependence of the blue fraction
may be the result of a decreasing
infall rate with time, and/or reflect the well established
overall decrease of
the star formation rate in the field population
(e.g., Kauffmann 1995).

It is usually assumed that a sizable
fraction of the blue population are progenitors of
(low mass) red early-type galaxies in nearby clusters
(e.g., Kodama \& Bower 2001), in apparent conflict with the
early formation of early-type galaxies inferred
from studies of their color and luminosity
evolution.

\subsection{Evolution of the early-type galaxy fraction}

Dressler et al.\ (1997) report a high fraction of spiral galaxies in clusters
at $0.3<z<0.5$. These galaxies are much rarer in nearby rich clusters,
and hence must have transformed into early-type galaxies between $z=0.5$
and $z=0$. Other studies (e.g., Couch et al.\ 1998, van Dokkum et al.\ 2001)
have confirmed this trend, and extended it to $z=1.3$. The evolution
of the early-type galaxy fraction is shown in Fig.\ 4. The early-type
fraction decreases by a factor $\sim 2$ from $z=0$ to $z \sim 1$, although
the trend has significant scatter.

\begin{figure}[h]
\plotone[-103 187 661 676]{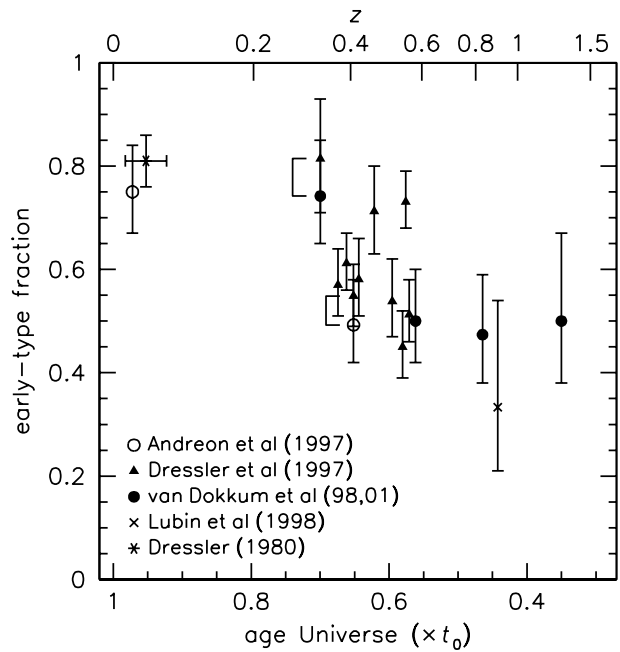}
\caption{Evolution of the early-type galaxy fraction compiled from
various studies, and taken from
van Dokkum et al.\ (2001).}
\end{figure}

Dressler et al.\ found that the increased fraction of spiral galaxies
at high redshift is accompanied by a low fraction of S0 galaxies,
and concluded that the $z\approx 0.4$ spiral galaxies transform
into S0 galaxies. They postulate that the formation of
elliptical galaxies predated the virialization of the clusters
in which they now live. However,
there is some controversy over the relative numbers
of elliptical and S0 galaxies in distant clusters\footnote{Importantly,
there is no such controversy over the {\em combined} number of Es and
S0s, i.e., the early-type galaxy fraction.}. Dressler et al.
claim that S0 galaxies are virtually absent at $z\sim 0.4$, suggesting
a factor $\sim 4$ evolution over the past $\sim 4$\,Gyr.  Others
have debated this, and find a much milder evolution in the E/S0 ratio
(e.g., Andreon et al.\ 1997).

One of the problems is the difficulty in
distinguishing elliptical and S0 galaxies at high redshift
(see Fabricant et al.\ 2000
and references therein). 
Spatially resolved kinematics
may offer an elegant solution.
An example is shown in Fig.\ 5. From the HST image alone
it is difficult to determine
whether this $z=0.83$ galaxy is an elliptical or an S0. The kinematics
reveal a rapidly rotating cold disk in addition to a hot bulge, and
show that this object is a massive S0.

\begin{figure}[h]
\plottwo{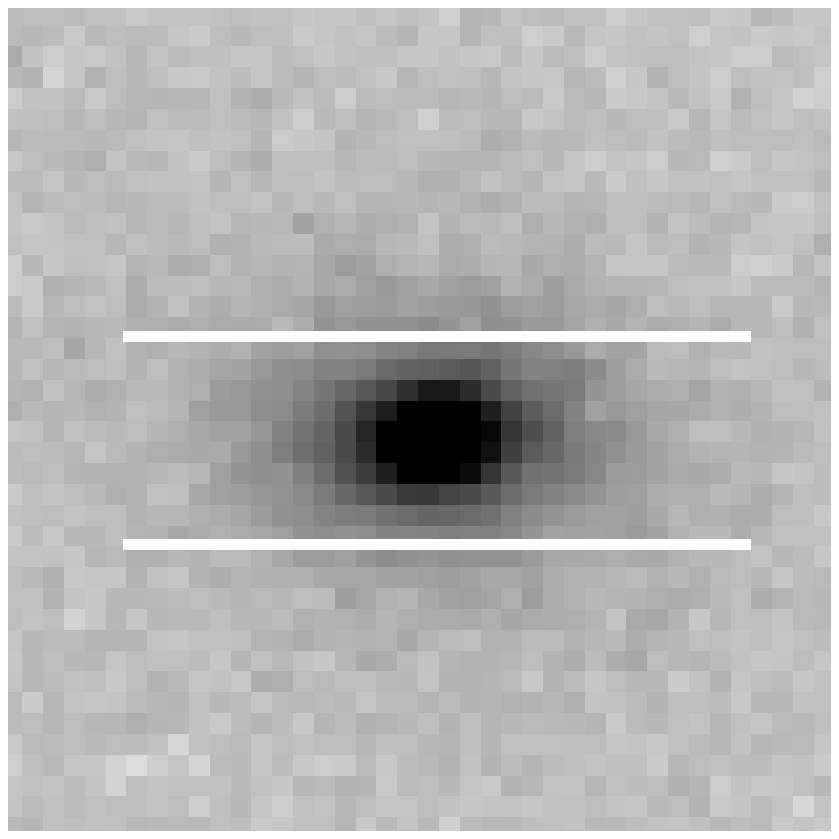}{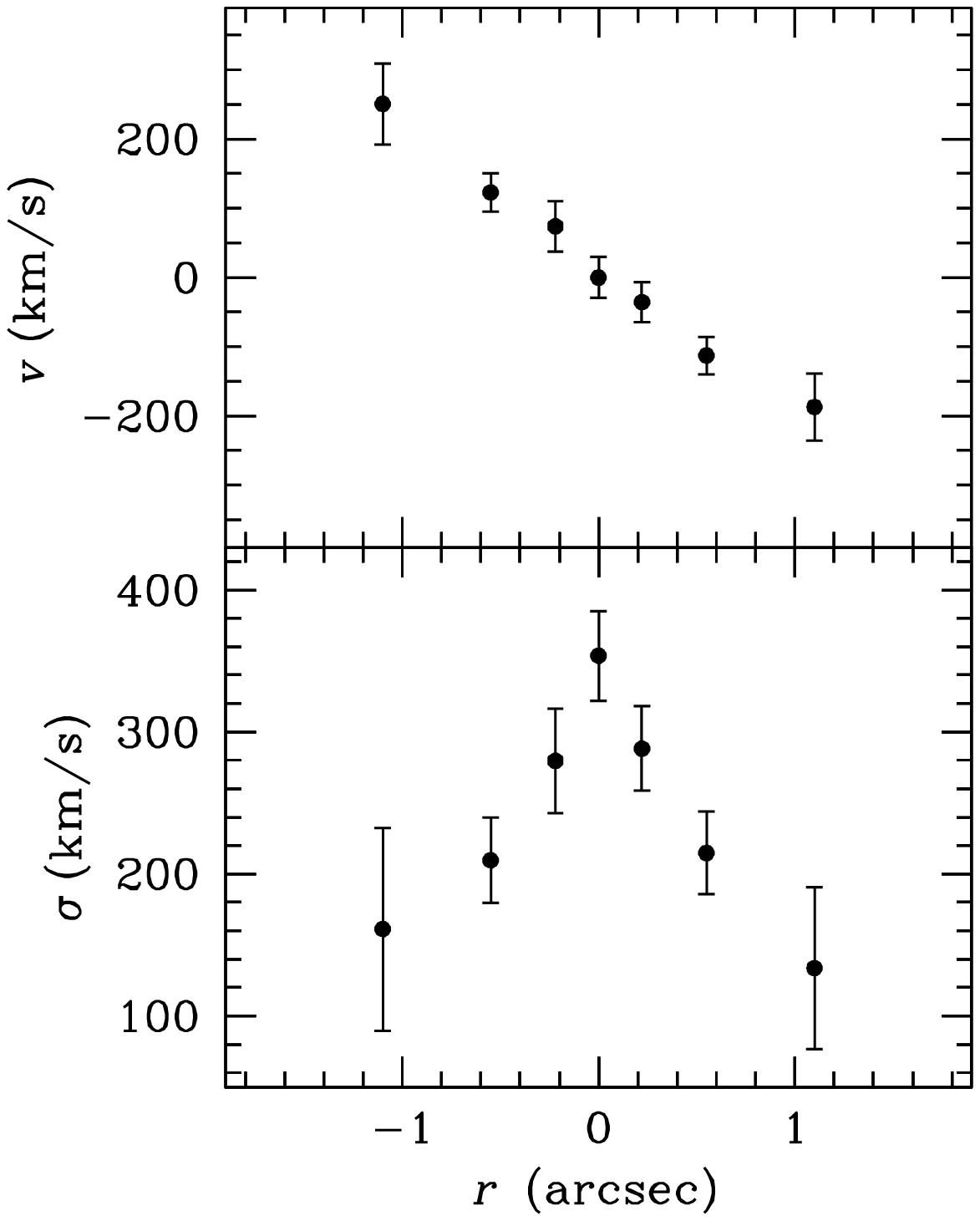}
\caption{
Based on the HST image shown at left
it is difficult to determine whether this $z=0.83$ galaxy is an
E or an S0. The kinematics reveal a
rapidly rotating disk, demonstrating that this is an S0.}
\end{figure}

\subsection{Mergers in $z\sim 1$ clusters}

The discovery of a large number of red merger systems in the cluster
MS\,1054--03 at $z=0.83$ is arguably the most spectacular evidence
for recent formation of massive early-type galaxies (van Dokkum
et al.\ 1999). We obtained deep, multi-color images of
this cluster at 6 pointings with WFPC2 on HST. Redshifts of
galaxies in this field were obtained with the Keck Telescope;
89 of those are cluster members. The survey is described in
van Dokkum et al.\ (2000).

We found that 17\,\% of the galaxies in MS\,1054--03 are merger
systems. Most of the mergers are very luminous ($M_B \sim -22$
in the rest frame, or $\sim 2 L_*$ at $z=0.83$), and a striking
way to display our result is to show a panel with the 16 brightest
confirmed cluster members (Fig.\ 6). Five were
classified as mergers.

The mergers are generally red, with a few exceptions. Similarly,
the spectra of most of the mergers do not show strong emission
lines. These results suggest that the bulk of the stars 
was formed well before the merger. Hence the stellar age of
the merged galaxies will be significantly different from
the ``assembly age''. The physical reason for the low star formation
is unknown: it is possible that the massive precursor galaxies
had already lost their cold gas due to internal processes
(such as super winds, or winds driven by nuclear activity).
Alternatively, the cold gas may have been stripped by the
cluster X-ray gas, or exhausted in an earlier phase of merging
in dense, infalling groups.

\begin{figure}
\plotone[82 207 530 556]{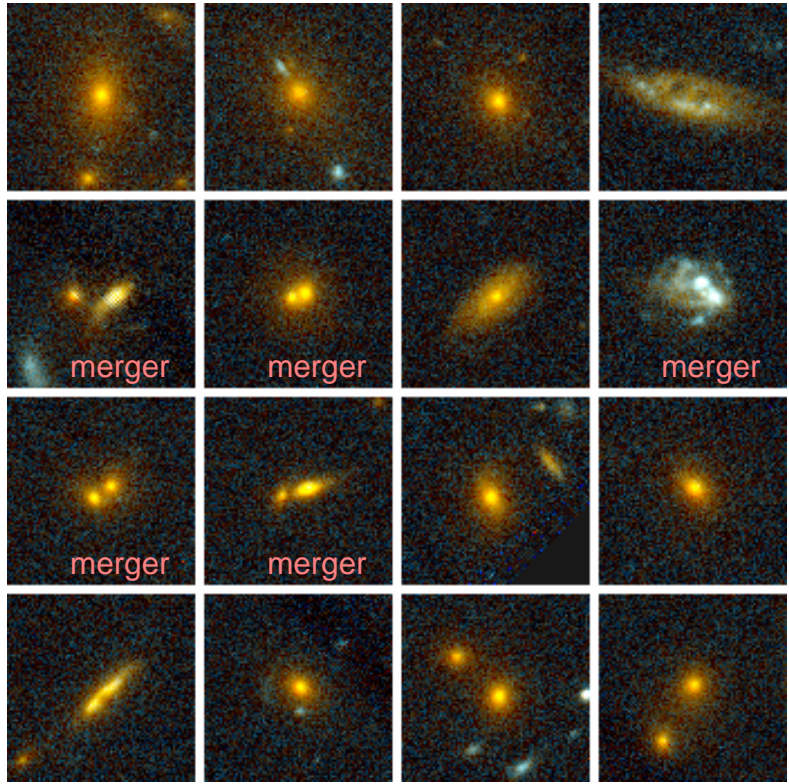}
\caption{The 16 brightest confirmed members of MS\,1054--03
at $z=0.83$, ordered by $I$ magnitude. Note the large number
of mergers. A color version of this figure can be found at
http://www.astro.caltech.edu/\~{ }pgd/ms1054.}
\end{figure}

It is not yet known whether the galaxy population of
MS\,1054--03 is typical for its redshift. One possibility is that such a phase
of enhanced merging occurs at different redshifts for different clusters.
In MS\,1054--03 the mergers probably occur in infalling subclumps,
and its high merger fraction could be related to its overall
unvirialized state (van Dokkum et al.\ 1999).

We recently completed a morphological study of an even higher
redshift cluster, \cl1\ at $z=1.27$ (van Dokkum et al.\ 2001).
The Brightest Cluster Galaxy
has an asymmetric outer envelope, demonstrating that it recently
experienced a merger or strong tidal interaction. The second
brightest galaxy is yet another red merger system, in this case
between three galaxies of comparable brightness. This remarkable
system shows that red mergers are not unique to MS\,1054--03,
and may even be common at $z\sim 1$. Studies of more clusters are
required to better quantify the role of these red mergers in the
formation of massive galaxies.

\section{Effects of morphological evolution: the progenitor bias}

The existence of
the Butcher-Oemler effect, the evolution of the early-type galaxy
fraction, and the presence of mergers in distant clusters
imply that simple models for the evolution of early-type galaxies
in clusters are insufficient. Therefore, we need to consider
more complex models that incorporate morphological transformations.
The problem was first addressed in Franx \& van Dokkum (1996), and worked
out in van Dokkum \& Franx (2001).

\subsection{Complex models}

We assume that early-type galaxies have two phases in their history:
 first a relatively long phase in
which they were forming stars, at a rate which can be constant, or
variable with time. Second, they are transformed into galaxies without
star formation, through a merger, gas-stripping, or other mechanism. 
During a period of $\approx$ 1 Gyr they are classified as
post-starburst galaxy, merger galaxy, or other ``special
type''. After that,  they are classified as normal early-type
galaxies.  As a result, the set of early-type galaxies
evolves, and the set of galaxies classified as early-types at
$z=0$ is not the same as the set of galaxies classified as
early-types at high redshift.

The evolution of the $M/L$ ratio of individual galaxies
is shown in Fig.\ 7c.
The evolution of the $M/L$ ratio of each galaxy is indicated with
a dotted line when it is not yet classified as
early-type, and with a continuous curve when it is
classified as early-type galaxy. As is obvious from the plot, the
continuous addition of young early-type galaxies to the sample has a
significant effect on the evolution of the mean $M/L$ ratio: the
newly added galaxies pull the mean $M/L$ ratio to lower values.
As a result the evolution of the mean $M/L$ ratio of the full
sample is very slow -- much slower than the evolution of the
$M/L$ ratio of ``typical'' individual galaxies. This
effect is called ``progenitor bias'': as we compare the $M/L$ ratios
of early-type galaxies at different redshifts, we compare between
different sets of galaxies, and the evolution of their properties
can be misinterpreted when morphological evolution is ignored.

The evolution of the mean $M/L$ ratio of the early-types is shown in
Fig.\ 7d. As can be seen, the mean evolution is slow. The slope of the
$M/L-z$ relation is comparable to the slope for a single galaxy which
formed at very high redshift, even though the mean formation redshift
of all early-type galaxies at $z=0$ is low at $z_{\rm form}=2$.
The long dashed curve in the figure
indicates the evolution of the mean $M/L$ ratio of all galaxies
classified as early-types at $z=0$. The evolution is much faster, as
expected. The difference between these two curves is  caused by the
progenitor bias, and it can be quite substantial.

The scatter for the early-types at any redshift is indicated by the
shaded region in Fig.\ 7d.
Because the youngest galaxies drop out of the
sample at higher redshift, the scatter remains constant,
or even decreases slightly at higher redshift.
The same holds for the scatter in the CM relation.
This counter-intuitive result can be explained by the fact that the
models are approximately scale free in time, and the
relative age differences between the early-type galaxies are
similar at all times.

\subsection{Application to data}

Figure 8 shows two models which fit the evolution of the $M/L$ ratio
measured from the Fundamental Plane, and the evolution of the scatter
in the color-magnitude relation. The morphological transformations
are described by a simple function which provides a good fit (panel a).
Two models are explored for the star formation rate during the
phase when the galaxies are spirals:
the solid  line indicates a model with constant star formation rate,
and the dotted line shows a model with declining star formation rate. 
They both fit the evolution of the
$M/L$ ratio well (panel b). The model with the declining star formation ratio
underpredicts the scatter in the color magnitude relation and $M/L$ 
ratios. Hence for
this model the scatter is not produced entirely by age differences,
but also by scatter in the metallicity-magnitude relation, or other effects.

\begin{figure}
\plotone[-28 347 430 689]{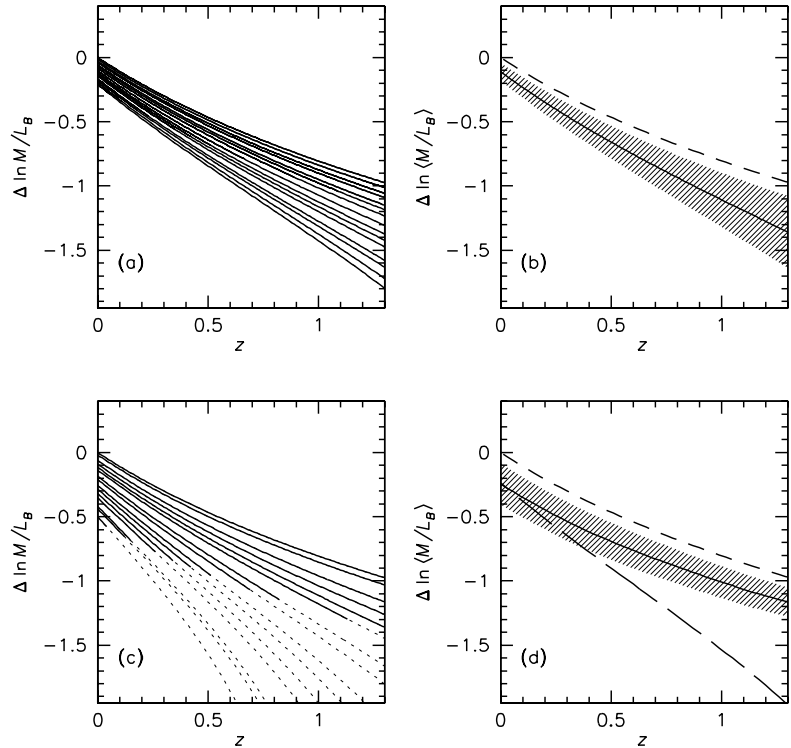}
\caption{Model predictions for the evolution of the $M/L$ ratio for
simple models (a,b), and complex models with morphological
transformations (c,d). See text and van Dokkum \& Franx (2001).}
\end{figure}

The progenitor bias for the two models is different, as might be
expected from the difference in the predicted scatter due to age
variations.  When the evolution of the $M/L$ ratio is fitted with a
simple model without morphological transformations, we obtain a mean
formation redshift $\zform=6.5$. The complex model with constant
star formation produces an estimate of $\zform=3$.  The model with a
declining star formation rate produces $\zform=4$.  These values
apply for a cosmology with $\Omega_m=0.3$. They change to 2.6, 2.0,
and 2.2, respectively, for a flat universe with $\Omega_m=0.3$.  Hence the
effects of the progenitor bias are modest, but should not  be ignored.

\begin{figure}
\plotone{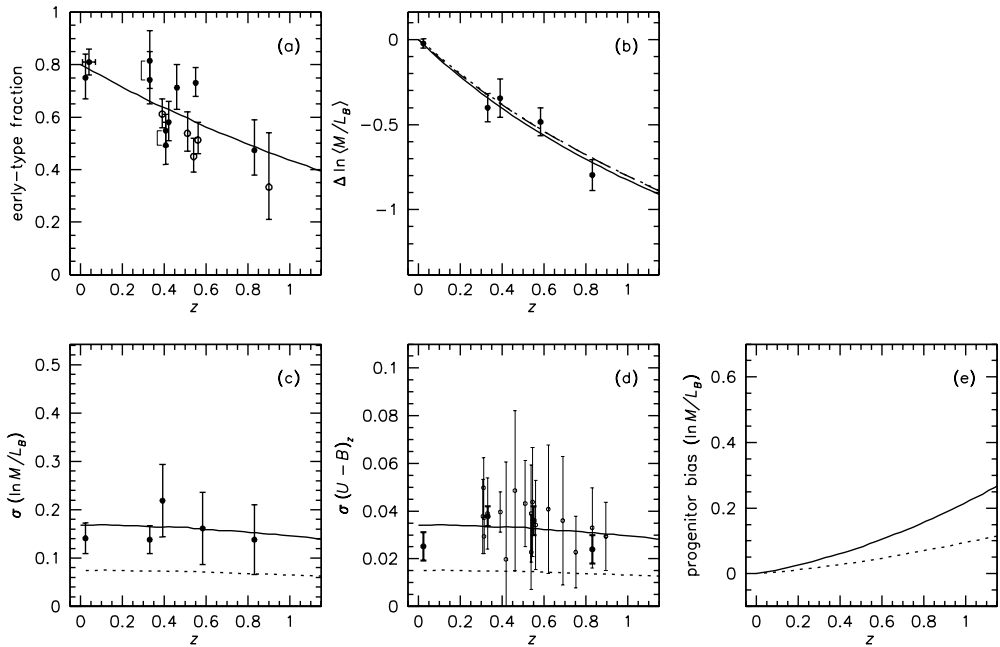}
\caption{Complex models fitted to observations.
Models which include morphological transformations can fit
the evolution of the early-type galaxy fraction (a), the mean $M/L$ ratio
(b),
and the scatter in the FP (c) and the CM relation (d) simultaneously.
The error that is made when using simple models (i.e., without morphological
transformations) is shown in (e). See van Dokkum \& Franx (2001)
for details.
}
\end{figure}

\section{Conclusions}

Models which include morphological transformations
can reconcile two apparently contradictory lines of evidence: the low
scatter in the color magnitude relation and the slow evolution of the
$M/L$ ratio on one hand, and the morphological  evolution observed
in rich clusters on the other. A basic framework for
galaxy evolution in clusters, which includes
infall and morphological transformations, seems to be developing
and future observations will be aimed at refining and
testing these ideas.
The available data at $z\sim 1$ are still sparse, and
the clusters that have been studied so far may not be typical
progenitors of ``run of the mill'' nearby clusters. It is expected
that ACS on HST will make it much easier to study clusters at
this epoch; the future of this field can therefore be considered bright!
\vspace{0.2cm}

It is a pleasure to thank the organizers for an interesting and lively
meeting in beautiful Sesto Pusteria, and for financial support.

\end{document}